\documentclass[aps,prd,nofootinbib,amsmath,amssymb,twocolumn,10pt]{revtex4}

\pdfoutput=1

\usepackage{graphicx}
\usepackage{dcolumn}
\usepackage{bm}
\usepackage{amssymb}
\usepackage{latexsym}
\usepackage{booktabs}
\usepackage{amsmath}
\usepackage{multirow}
\usepackage[colorlinks=true, linkcolor=red, citecolor=blue]{hyperref}

\newcommand{\be}{\begin{equation}}
\newcommand{\ee}{\end{equation}}
\newcommand{\bq}{\begin{eqnarray}}
\newcommand{\eq}{\end{eqnarray}}

\usepackage[usenames,dvipsnames]{xcolor}

\begin{document}

\title{Gravitational wave standard sirens and cosmological parameter measurement}

\author{Xin Zhang}
\affiliation{Department of Physics, College of Sciences, Northeastern
University, Shenyang 110819, China} 
\affiliation{Ministry of Education Key Laboratory of Data Analytics and Optimization
for Smart Industry, Northeastern University, Shenyang 110819, China}
\affiliation{Center for High Energy Physics, Peking University, Beijing 100080, China}

\begin{abstract}


Gravitational wave standard sirens can be used to arbitrate the $H_0$ tension within the following several years. In the future, the observation of gravitational wave standard sirens would be developed into a powerful new cosmological probe because they can play an important role in breaking parameter degeneracies formed by other observations. Therefore, gravitational wave standard sirens are of great importance for the future accurate measurement of cosmological parameters.

\end{abstract}
\maketitle

The first detection of gravitational waves (GWs) produced by the binary neutron star (BNS) merger in August 17, 2017 \cite{TheLIGOScientific:2017qsa} is fairly meaningful because it initiated the new era of multi-messenger astronomy. In this BNS merger event, we not only detected the gravitational waves, but also detected the electromagnetic waves in various wave bands \cite{GBM:2017lvd}. This multi-messenger observation led to a series of important research progresses in the areas of astronomy and cosmology.

It is well-known that the GW observations can be used as cosmic standard sirens provided that the GW's waveform and the electromagnetic information of the event can both be obtained, which was first proposed by Schutz in 1986 \cite{Schutz1986} and was subsequently discussed by Holz and Hughes in 2005 \cite{Holz:2005df}. According to the Hubble's law, if we both accurately know the redshift and distance of an object, we can then directly derive the Hubble constant. Of course, the condition is that the redshift should be small enough. The multi-messenger observation of the BNS merger \cite{GBM:2017lvd} realized such an independent measurement of the Hubble constant, giving a result of $H_0=70.0^{+12.0}_{-8.0}$ km s$^{-1}$ Mpc$^{-1}$ \cite{Abbott:2017xzu}. The main advantage of this standard siren method is that it avoids using the cosmic distance ladder. But due to the fact that we have only one data point, the error is still large, around 15\%. 

\begin{figure}[!htp]
\includegraphics[scale=0.28]{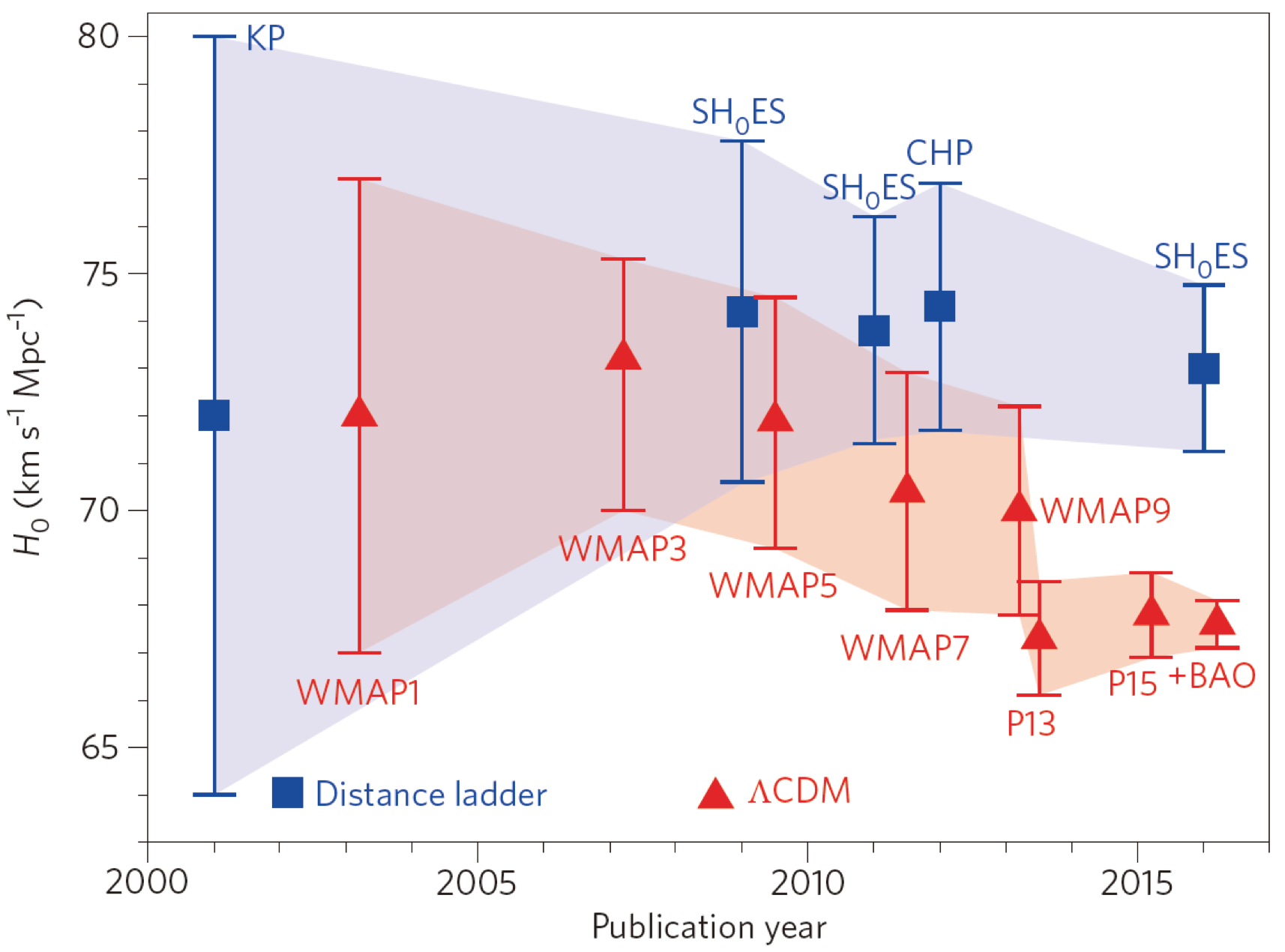}
\centering
\caption{\label{fig1} Tension in the determination of the Hubble constant $H_0$. Measurement values of $H_0$ as a function of publication year since the Hubble Key Project. Symbols in blue represent the values of $H_0$ determined in the nearby universe with a calibration based on the Cepheid distance scale. Symbols in red represent the derived values of $H_0$ based on an adopted cosmological model and measurements of CMB. Labels indicate the different experiments and data sets used for the determination of $H_0$ values. The blue- and red-shaded regions show the evolution of the uncertainties in these values, which have been decreasing for both methods. Currently, the two methods show a great tension between each other. This figure is taken from Ref.~\cite{Freedman:2017yms}.} 
\end{figure}

The Hubble constant $H_0$ is the first cosmological parameter, which was proposed in the 1920s, and so actually it has been measured for about one century. We have two methods to measure the Hubble constant: One is a model-dependent method using the cosmic microwave background (CMB) anisotropies measurement, such as the Planck data, combined with other astrophysical data, to constrain a cosmological model, and to obtain the fit result of the Hubble constant; so, this is a method of early universe. The other one is a model-independent method using the cosmic distance ladder to directly determine the Hubble constant; so, this is a method of late universe. Currently, the late-universe method has achieved the accuracy of 2\%, and the early-universe method has achieved the accuracy of 0.7\%. However, the problem is that the two results are now in great tension. The distance ladder gives a higher value of $H_0$, and the Planck observation favors a lower value of $H_0$, and they are in more than 4$\sigma$ tension \cite{Freedman:2017yms}. According to the latest result of distance ladder method by Riess et al. \cite{Riess:2019cxk}, the tension is now at the spectacular 4.4$\sigma$ level. See Fig.~\ref{fig1} for the current tension in the determination of $H_0$. Actually, the Hubble constant problem is one of the most important problems in the current cosmology. We do not know which one is correct, and so we say that cosmology is now at a crossroads \cite{Freedman:2017yms}.

\begin{figure}[!htp]
\includegraphics[scale=0.25]{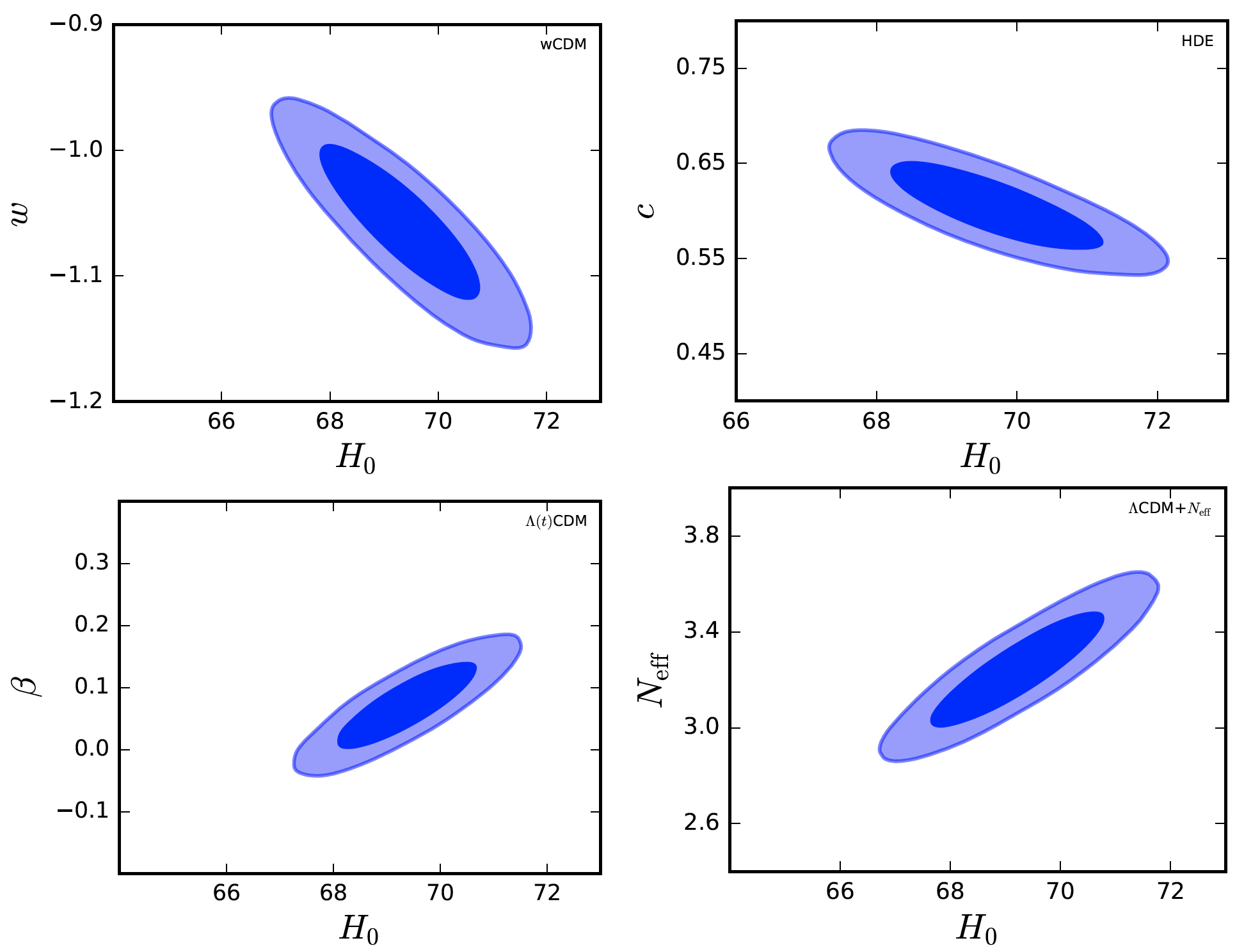}
\centering
\caption{\label{fig2} Two-dimensional marginalized contours in the $w$--$H_{0}$ plane for the $w$CDM model, in the $c$--$H_{0}$ plane for the HDE model, in the $\beta$--$H_{0}$ plane for the $\Lambda (t)$CDM model, and in the $N_{\rm eff}$--$H_{0}$ plane for the $\Lambda$CDM+$N_{\rm eff}$ model, by using the CMB+BAO+SN+$H_{0}$ data. This figure is taken from Ref.~\cite{Guo:2018ans}.}
\end{figure}


Actually, the $H_0$ tension may be a hint for the possibility that there is some new physics beyond the standard 6-parameter $\Lambda$ cold dark matter ($\Lambda$CDM) cosmology. Considering new physical effects needs to introduce new parameters into the extended models, which may be in some correlations with the Hubble constant. Indeed, in some extended models, the newly introduced parameters do positively correlate or anti-correlate with $H_0$. Several typical examples are shown in Fig.~\ref{fig2}. In the $w$CDM model where $w$ is assumed to be a constant, we can see that $w$ is in anti-correlation with $H_0$. This means that, according to the current observational constraints, a phantom energy actually favors a higher $H_0$. Likewise, in the holographic dark energy (HDE) model, the parameter $c$ is also in anticorrelation with $H_0$, indicating that a late-time phantom holographic dark energy favors a higher value of $H_0$. Next, let's see the situation of the interacting dark energy scenario (see Ref.~\cite{Zhang:2017ize} for a recent brief review for the related topic). 
In a specific case of interacting vacuum energy model, in which the interaction term is proportional to the cold dark matter density, we can see that the coupling parameter $\beta$ is in positive correlation with $H_0$, showing that an effective phantom also favors a higher value of $H_0$. At last, let's see the $\Lambda$CDM+$N_{\rm eff}$ model where the dark radiation is considered. We can see that $N_{\rm eff}$ is also in positive correlation with $H_0$, so a higher $N_{\rm eff}$ favors a higher $H_0$. But now the question is if the $H_0$ tension can really be resolved in these extensions to the $\Lambda$CDM cosmology. Actually, these models have been tightly constrained by the current observations. In a recent work \cite{Guo:2018ans} we make a comparison for them and we wish to see if there is an extended model that can truly resolve the $H_0$ tension. 

The fit results for the single-parameter and multi-parameter extensions are shown in Tables 1 and 2 of Ref.~\cite{Guo:2018ans}. We use the CMB+BAO+SN+$H_0$ data combination to constrain a series of extended models. In order to obtain a larger $H_0$ in the global fit, we use the $H_0$ data in the combined data sets. We find that, for the $\Lambda$CDM model, even though the $H_0$ data is used, the tension is still large, still at around the 3$\sigma$ level. Among these models, we can easily find that the HDE model and the HDE plus sterile neutrino model look the best ones in alleviating the $H_0$ tension, because for them the tension is at 1.67$\sigma$ and 1.11$\sigma$, respectively, rather small already. But, we should also notice that actually they have been not favored by the current observations. The $\chi^2_{\rm min}$ and Akaike information criterion  (AIC) values of them are very large. We find that the AIC values of them are larger than that of the $\Lambda$CDM model by about 20 or more, indicating that they have been excluded by the current observations. Thus, they should be removed from the list of considered extended models. Next, let's have a look at the $\Lambda$CDM+$N_{\rm eff}$ model, which now looks the best one among the remaining models, because the tension of it is at only the 1.87$\sigma$ level, and it is also favored by the current observations, i.e., the $\chi^2_{\rm min}$ and AIC values of it are similar to those of the $\Lambda$CDM model. But, we find that if the $H_0$ prior is not used in the data combination, then the tension immediately becomes at the 2.66$\sigma$ level. In addition, increasing $N_{\rm eff}$ can increase $\sigma_8$, which is another tension. Therefore, even though using the $N_{\rm eff}$ parameter can release the $H_0$ tension, it actually exacerbates the $\sigma_8$ tension. Through a comprehensive analysis, we can draw our conclusion \cite{Guo:2018ans}: that among these extensions, no one can truly resolve the $H_0$ tension.

Facing the two in-tension $H_0$ measurements, we actually need a third party to arbitrate the $H_0$ measurements. Actually, GW observations can provide such a third party. GW observations can arbitrate the $H_0$ measurements, because GWs can serve as standard sirens. The advantage of GW standard siren observations is that they provide a pure distance measurement, avoiding the complex astrophysical distance ladder and poorly understood calibration process. They are directly calibrated by theory. But, currently, we have only one data, so the error is still large, about 15\%, and so it cannot make an arbitration yet. In the future, we will have more low-redshift standard siren data, and so the error will be reduced to $15\%/\sqrt{N}$, with $N$ being the number of low-redshift standard siren data. Therefore, if we have 50 data, then the error will be decreased to about 2\% \cite{Chen:2017rfc}, similar to the error of the current distance ladder result. Actually, in the near future, the KAGRA and LIGO-India will join the detector network, and thus the error will become smaller, around $13\% /\sqrt{N}$ \cite{Chen:2017rfc}. Furthermore, in the future, we have a plan to build the third generation ground-based GW detectors, such as the Cosmic Explorer (CE) \cite{Evans:2016mbw} and the Einstein Telescope (ET) \cite{Punturo:2010zz}. Hence, we have confidence to judge which one is correct in the future by using the GW standard siren observations.

In Ref.~\cite{Chen:2017rfc}, the authors reported that a 2\% Hubble constant measurement from standard sirens will be achieved within five years. They say that in 2023, about 50 events will be observed by the Advanced LIGO-Virgo network and 2\% error will be achieved; in 2026, about 100 events will be observed by the five detectors and about 1\% error will be achieved. In the future, the third generation ground-based GW detectors will be built, e.g., the ET, which has 10 kilometer-long arms and three detectors. Obviously, the detection ability of ET is much better than the Advanced LIGO. Compared to the Advanced LIGO, ET has a much wider detection frequency range and a much better detection sensitivity. Thus, using the ET one can observe more BNS events in much deeper redshifts. A conservative estimation tells us that in a 10-year run of ET, about 1000 useful standard sirens can be observed \cite{Zhang:2018byx}. Therefore, in the future, it is expected that the standard sirens would be developed into a powerful cosmological probe.

\begin{figure}[!htp]
\includegraphics[scale=0.3]{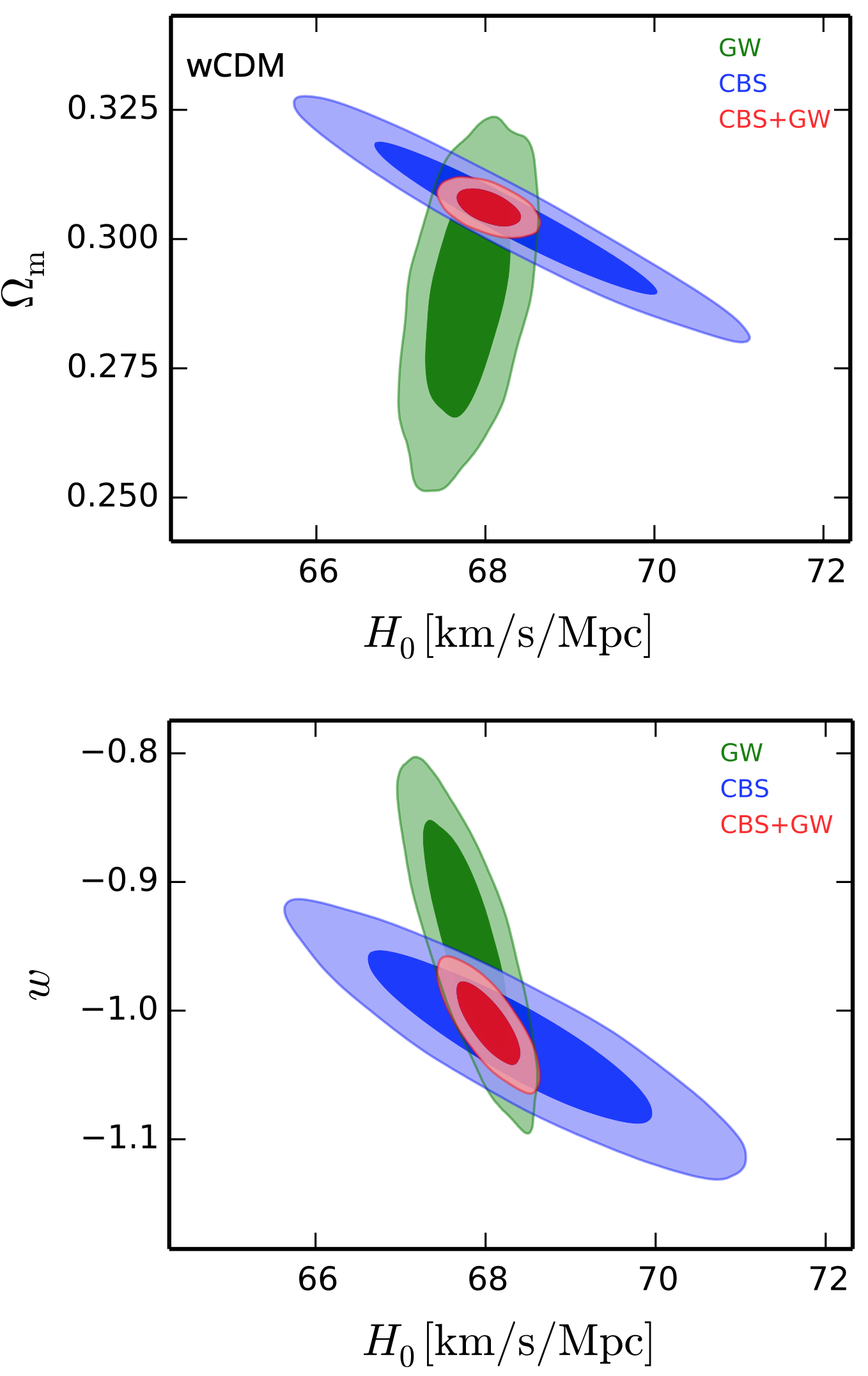}
\centering
\caption{\label{fig3} Constraints (68.3\% and 95.4\% confidence level) on the $w$CDM model in the $\Omega_{\rm m}$--$H_0$ plane ({\it upper}) and in the $w$--$H_0$ plane ({\it lower}) using GW, CBS, and CBS+GW data combinations. Here, CBS stands for CMB+BAO+SN. This figure is taken and adapted from Ref.~\cite{Zhang:2018byx}.}
\end{figure}

It is therefore rather necessary to study what role the standard sirens would play in the cosmological parameter estimation in the future. In a recent work \cite{Zhang:2018byx}, we take the ET as an example to make an analysis. We find that the standard sirens are very good at measuring the Hubble constant, but for the measurements of other cosmological parameters they are not so good. In Ref.~\cite{Zhang:2018byx}, we show that the measurement of $H_0$ by GW alone is at a 0.3\% precision for $\Lambda$CDM and at a 0.5\% precision for $w$CDM. Because the standard sirens can measure the absolute luminosity distance, they can break the parameter degeneracies formed by other observations. Therefore, the standard sirens are very meaningful in this sense for the cosmological parameter measurements. In the upper panel of Fig.~\ref{fig3}, we can see that for the $w$CDM model the contours in the $\Omega_{\rm m}$--$H_0$ plane by GW alone and CMB+BAO+SN are roughly orthogonal, and thus the degeneracy is broken thoroughly. In the lower panel of Fig.~\ref{fig3}, we can see that the GW standard siren observation cannot provide a good enough measurement for $w$, with the precision being only 6\%; as a contrast, the current CMB+BAO+SN observation can give an about 4\% measurement for $w$. But due to the fact that the degeneracy formed by the current CMB+BAO+SN observation can be broken by the future GW observation, the combined data of them will give a 2\% precision for the $w$ measurement. Therefore, we expect that in the future the GW observation combined with other future survey projects would be capable of elucidating the nature of dark energy.

In Ref.~\cite{Wang:2018lun}, we show that the future GW standard sirens observed by ET can also improve the constraints on the neutrino mass by about 10\%. For a recent minor review for the topic of weighing neutrinos in cosmology, see Ref.~\cite{Zhang:2017rbg}. 
Furthermore, we also check for some other dark energy models, and we find that the standard sirens can play an important role in breaking parameter degeneracies for all the considered models (see, e.g., Refs.~\cite{Zhang:2019ple,gwde}).

By the way, here we also mention that the result of gravitational wave propagation speed exactly equal to the speed of light, at the precision of $10^{-15}$, inferred from the BNS merger event (GW170817 and GRB 170817A) \cite{Monitor:2017mdv}, has exerted a surprisingly impact on the research field of dark energy. A significant fraction of the parameter space of theories involving a scalar field coupled to gravity, which leads to the gravitational wave propagation speed varied, has been ruled out by this experimental result \cite{Baker:2017hug,Creminelli:2017sry,Sakstein:2017xjx,Ezquiaga:2017ekz}.

We have entered the precision cosmology era. Great achievements have been made in the study of cosmology. For example, some cosmological parameters have been precisely measured; the acceleration of the expansion of the universe has been discovered; and, a cosmological standard model has been established. But the question is if we believe that using only six parameters can entirely describe the evolution of the universe. Actually, most of us do not believe. The current situation actually indicates that the observations are still not accurate enough, and they cannot accurately measure the other parameters beyond the standard model. The present main problems in the cosmological parameter measurement are the following two: (i) There are inconsistencies between some observations, and (ii) there are degeneracies between some parameters. These problems indicate that cosmological models should be further extended, and cosmological probes should also be further developed. 

The current mainstream mature cosmological probes mainly include: CMB anisotropies, type Ia supernovae, baryon acoustic oscillations (BAO), the Hubble constant $H_0$ determination, weak lensing, clusters of galaxies, and redshift-space distortions (RSD). In the future, these probes will be further developed. For example, we have the fourth generation dark energy programs such as LSST \cite{Abell:2009aa}, Euclid \cite{Laureijs:2011gra} and WFIRST \cite{Spergel:2013tha}. But, we also need other new cosmological probes other than the optical survey programs. In my opinion, the most important new cosmological probes are the radio observation and the GW observation. The future neutral hydrogen 21-centimeter survey by SKA \cite{Bacon:2018dui} would provide neutral hydrogen power spectrum, and related BAO and RSD signals. In addition, the future GW observations by ground-based and space-based detectors will be used as standard sirens. So, in the future 10 to 15 years, we will have some powerful cosmological probes to study dark energy and to precisely measure cosmological parameters. The GW standard siren observation combined with the optical, near-infrared, and radio survey observation programs (e.g., LSST, Euclid, WFIRST, SKA, and so forth) will greatly promote the development of cosmology.

At last, a brief summary is given. Binary neutron star collision observations opened a new era to multi-messenger astronomy. Gravitational wave standard sirens do not depend on the distance ladder, and they can measure the absolute cosmological distance. The tension in the Hubble constant measurements is very important in the current cosmology, and it seems that the extended cosmological models cannot resolve the tension problem. Future gravitational wave observations may arbitrate the $H_0$ tension. Standard sirens will be developed into a powerful new cosmological probe in the future, because they can play an important role in breaking parameter degeneracies. Future gravitational wave observations combined with other optical surveys perhaps can elucidate the nature of dark energy.

\begin{acknowledgments}
This work was supported by the National Natural Science Foundation of China (Grants Nos.~11835009, 11690021, and 11522540) and the National Program for Support of Top-Notch Young Professionals.


\end{acknowledgments}



\end{document}